\documentclass[preprint,final, nomark]{elsarticle}
\makeatletter
\def\ps@pprintTitle{%
 \let\@oddhead\@empty
 \let\@evenhead\@empty
 \let\@oddfoot\@empty
 \let\@evenfoot\@empty
}
\makeatother

\usepackage{graphicx} 
\usepackage{amsmath,amssymb}
\usepackage{enumerate}
\usepackage{hyperref} 

\usepackage[T1]{fontenc}
\usepackage{lineno,hyperref}
\usepackage{upgreek}
\usepackage{xcolor}
\usepackage{lipsum}
\usepackage{orcidlink}

\newcommand*{\Apr}{{A^\prime}}

\begin{document}

\begin{frontmatter}

\begin{figure}[h!]
\begin{minipage}{\textwidth}
\centering
\includegraphics[width=0.5\textwidth, angle=0]{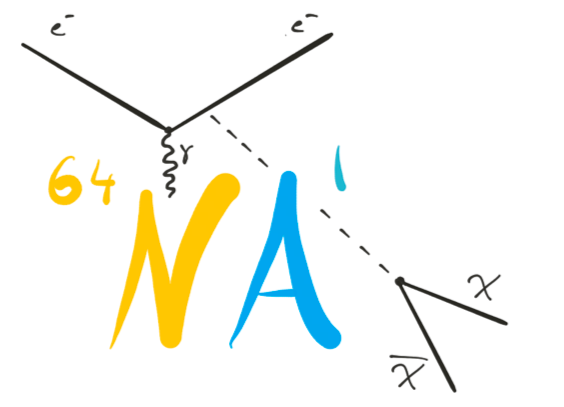}
\end{minipage}
\end{figure}

\title{Searching for Light Dark Matter and Dark Sectors with the NA64 experiment at the CERN SPS\footnote{Submitted as input to the European Particle Physics Strategy Update 2025.\\Edited by \href{P. Crivelli}{paolo.crivelli@cern.ch}, \href{S. Gninenko}{sergei.gninenko@cern.ch}, \href{L. Molina Bueno}{laura.molina.bueno@cern.ch}, \href{A. Celentano}{andrea.celentano@cern.ch}, \href{V. Poliakov}{vladimir.poliakov@cern.ch}}}

\author[a]{Yu.~M.~Andreev\orcidlink{0000-0002-7397-9665}}
\author[b]{A.~Antonov\orcidlink{0000-0003-1238-5158}}
\author[c,d]{M.~A.~Ayala~Torres\orcidlink{0000-0002-4296-9464}}
\author[e]{D.~Banerjee\orcidlink{0000-0003-0531-1679}}
\author[f]{B.~Banto Oberhauser\orcidlink{0009-0006-4795-1008}}
\author[g]{V.~Bautin\orcidlink{0000-0002-5283-6059}}
\author[e]{J.~Bernhard\orcidlink{0000-0001-9256-971X}}
\author[b,h]{P.~Bisio\orcidlink{/0009-0006-8677-7495}}
\author[b]{A.~Celentano\orcidlink{0000-0002-7104-2983}}
\author[e]{N.~Charitonidis\orcidlink{0000-0001-9506-1022}}
\author[f]{P.~Crivelli\orcidlink{0000-0001-5430-9394}}
\author[a]{A.~V.~Dermenev\orcidlink{0000-0001-5619-376X}}
\author[a]{S.~V.~Donskov\orcidlink{0000-0002-3988-7687}}
\author[a]{R.~R.~Dusaev\orcidlink{0000-0002-6147-8038}}
\author[g]{V.~N.~Frolov}
\author[g]{S.~V.~Gertsenberger\orcidlink{0009-0006-1640-9443}}
\author[e]{S.~Girod}
\author[a]{S.~N.~Gninenko\orcidlink{0000-0001-6495-7619}}
\author[g]{A.V.Ivanov}
\author[g]{Y.~Kambar\orcidlink{0009-0000-9185-2353}}
\author[a]{A.~E.~Karneyeu\orcidlink{0000-0001-9983-1004}}
\author[g]{G.~Kekelidze\orcidlink{0000-0002-5393-9199}}
\author[l]{B.~Ketzer\orcidlink{0000-0002-3493-3891}}
\author[a]{D.~V.~Kirpichnikov\orcidlink{0000-0002-7177-077X}}
\author[a]{M.~M.~Kirsanov\orcidlink{0000-0002-8879-6538}}
\author[a,g]{V.~A.~Kramarenko\orcidlink{0000-0002-8625-5586}}
\author[a,g]{N.~V.~Krasnikov\orcidlink{0000-0002-8717-6492}}
\author[c,d]{S.~V.~Kuleshov\orcidlink{0000-0002-3065-326X}}
\author[d]{V.~E.~Lyubovitskij\orcidlink{0000-0001-7467-572X}}
\author[b]{A.~Marini\orcidlink{0000-0002-6778-2161}}
\author[b]{L.~Marsicano\orcidlink{0000-0002-8931-7498}}
\author[g]{V.~A.~Matveev\orcidlink{0000-0002-2745-5908}}
\author[d]{R.~Mena~Fredes}
\author[d,m]{R.~Mena~Yanssen}
\author[n]{L.~Molina Bueno\orcidlink{0000-0001-9720-9764}}
\author[f]{M.~Mongillo\orcidlink{0009-0000-7331-4076}}
\author[g]{D.~V.~Peshekhonov\orcidlink{0009-0008-9018-5884}}
\author[a]{V.~A.~Polyakov\orcidlink{0000-0001-5989-0990}}
\author[o]{B.~Radics\orcidlink{0000-0002-8978-1725}}
\author[n]{K.~Salamatin\orcidlink{0000-0001-6287-8685}}
\author[a]{V.~D.~Samoylenko}
\author[f]{H.~Sieber\orcidlink{0000-0003-1476-4258}}
\author[a]{D.~Shchukin\orcidlink{0009-0007-5508-3615}}
\author[d,p]{O.~Soto}
\author[a]{V.~O.~Tikhomirov\orcidlink{0000-0002-9634-0581}}
\author[a]{I.~Tlisova\orcidlink{0000-0003-1552-2015}}
\author[a]{A.~N.~Toropin\orcidlink{0000-0002-2106-4041}}
\author[n]{M.~Tuzi\orcidlink{0009-0000-6276-1401}}
\author[a,g]{P.~V.~Volkov\orcidlink{0000-0002-7668-3691}}
\author[a]{I.~V.~Voronchikhin\orcidlink{0000-0003-3037-636X}}
\author[c,d]{J.~Zamora-Sa\'a\orcidlink{0000-0002-5030-7516}}
\author[g]{A.~S.~Zhevlakov\orcidlink{0000-0002-7775-5917}}

\affiliation[a]{organization={Authors affiliated with an institute covered by a cooperation agreement with CERN}}
\affiliation[b]{organization={INFN, Sezione di Genova}, postcode={16147}, city={Genova}, country={Italia}}
\affiliation[c]{organization={Center for Theoretical and Experimental Particle Physics, Facultad de Ciencias Exactas, Universidad Andres Bello}, city={Fernandez Concha 700, Santiago}, country={Chile}}
\affiliation[d]{organization={Millennium Institute for Subatomic Physics at High-Energy Frontier (SAPHIR)}, city={Fernandez Concha 700, Santiago}, country={Chile}}
\affiliation[e]{organization={CERN, European Organization for Nuclear Research}, postcode={CH-1211}, city={Geneva}, country={Switzerland}}
\affiliation[f]{organization={ETH Z\"urich, Institute for Particle Physics and Astrophysics}, postcode={CH-8093},city={Z\"urich},country={Switzerland}}
\affiliation[g]{organization={Authors affiliated with an international laboratory covered by a cooperation agreement with CERN}}
\affiliation[h]{organization={Universit\`a degli Studi di Genova}, postcode={16126}, city={Genova}, country={Italia}}
\affiliation[i]{organization={INFN, Sezione di Catania}, postcode={95123},city={Catania}, country={Italia}}
\affiliation[l]{organization={Universit\"{a}t Bonn, Helmholtz-Institut f\"ur Strahlen-und Kernphysik}, postcode={53115}, city={Bonn}, country={Germany}}
\affiliation[m]{organization={Universidad T\'ecnica Federico Santa Mar\'ia and CCTVal}, postcode={2390123}, city={Valpara\'iso}, country={Chile}}
\affiliation[n]{organization={Instituto de Fisica Corpuscular (CSIC/UV)}, city={Carrer del Catedratic Jose Beltran Martinez, 2, 46980 Paterna, Valencia}, country={Spain}}
\affiliation[o]{organization={Department of Physics and Astronomy, York University}, city={Toronto, ON}, country={Canada}}
\affiliation[p]{organization={Departamento de Fisica, Facultad de Ciencias, Universidad de La Serena}, city={Avenida Cisternas 1200, La Serena}, country={Chile}}



\begin{abstract}
Since its approval in 2016, NA64 has pioneered LDM searches with electron \cite{NA64:2023wbi}, positron \cite{NA64:2023ehh}, muon \cite{NA64:2024klw}, and hadron \cite{NA64:2024azv} beams. The experiment has successfully met its primary objectives, as outlined in the EPPS input (2018), and even exceed them producing results that demonstrate its ability to operate in a near-background-free environment. The Physics Beyond Collider (PBC) initiative at CERN recognize NA64’s contributions as complementary and worthy of continued exploration. Its key advantage over beam-dump approaches is that the signal rate scales as $(\text{coupling})^2$ rather than $(\text{coupling})^4$, reducing the required beam particles for the same sensitivity.

To fully exploit the NA64 physics potential, an upgrade during LS3 will enable NA64 to run in background-free mode at higher SPS beam rates. Planned upgrades include (a) improved detector hermeticity with a new veto hadron calorimeter, (b) enhanced particle identification with a synchrotron radiation detector, and (c) increased beam rates via upgraded electronics. 

With the recently strengthened NA64 collaboration, stable operations and timely data analysis are planned for LHC Run 4. The expected $\sim 10^{13}$ electrons, $\sim 10^{11}$ positrons (40 and 60 GeV), and $\sim 2\times10^{13}$ muons on target will allow NA64 to explore new light dark matter regions, with the potential for discovery or conclusive exclusion of many well-motivated LDM models.
\end{abstract}

\end{frontmatter}

\section{Introduction}

Despite intensive searches at the LHC and in non-accelerator experiments, Dark Matter remains a mystery that challenges our understanding of the Universe (see, e.g., \cite{RevModPhys.90.045002} for a recent review). One of the most popular and extensively searched-for thermally produced DM candidates falls under the category of Weakly Interacting Massive Particles (WIMPs), a term that encompasses the lightest supersymmetric particles, Kaluza-Klein states from extra-dimensional models and others. The negative results from the extensive WIMP detection program are pushing further investigations toward the high-energy and high-sensitivity frontiers (see, e.g., \cite{Feng:2010gw} and references therein).

In recent years, a variety of alternative DM candidates have been proposed to address this fundamental question. A broad class of well-motivated models introduces the concept of \textbf{"Dark" (or hidden) Sectors} (DS), providing a natural framework to explain the origin and properties of DM. In these models, DM is part of a hidden sector composed of particles that are singlets under the $SU(3)_C \times SU(2)_L \times U(1)_Y$ gauge group of the Standard Model (SM), interacting with ordinary matter through gravity and potentially other feeble forces.

From an experimental perspective, the sensitivity of searches for these new singlet states depends on the coupling strength and mass scale of the dark particles. To probe the parameter space of the most motivated DS theories, it is essential to employ a combination of search techniques optimized for different mass ranges and interaction strengths.

One of the most intriguing scenarios postulates the existence of a new portal interaction between the DS and visible matter, mediated by a new vector boson $A'$, the so-called \textit{dark photon}. The $A'$ could be light, with a mass $m_{A'} \lesssim 1$ GeV, associated with a spontaneously broken $U(1)_D$ gauge symmetry, and could interact with the SM through kinetic mixing with the ordinary photon. This interaction is described by the additional term $-\frac{1}{2} \epsilon F_{\mu\nu}F'^{\mu\nu}$ in the massive photon Lagrangian, parameterized by a small mixing strength $\epsilon \ll 1$. A dark photon mass in the sub-GeV range can arise in various physics scenarios. If the additional secluded $U(1)_D$ symmetry is embedded in a Grand Unified Theory (GUT), the mixing can be generated at the one- or two-loop level, naturally yielding values of $\epsilon \simeq 10^{-3} - 10^{-1}$ (one loop) or $\epsilon \simeq 10^{-5} - 10^{-3}$ (two loops) \cite{ArkaniHamed:2008qn, Holdom:1985ag, Okun:1982xi}. The concept of sterile photons was first introduced by Okun in his paraphoton model \cite{Okun:1982xi} (see also \cite{Holdom:1985ag}).

One of the strongest motivations for the existence of light Dark Sectors is that they offer a viable framework to explain DM as a thermal freeze-out relic in a broader and lower mass range compared to WIMPs \cite{Feng:2008ya}. In Light Dark Matter (LDM) models, the dark state can account for the observed DM relic density \cite{deNiverville:2011it, Izaguirre:2015yja} through the so-called ``WIMPless miracle.'' This possibility has sparked a worldwide theoretical and experimental effort to search for dark forces and other portals connecting the visible and hidden sectors. As a result, the focus has shifted from high-energy colliders to the high-intensity frontier (see, e.g., \cite{Jaeckel:2010ni, Beacham:2019nyx, Fabbrichesi:2020wbt,Antel:2023hkf} for a review). DS could also provide a solution to other open questions of the SM such as the origin of the neutrino mass and the strong CP problem. 

\section{Brief history of NA64}

The NA64 experiment at CERN's SPS was primarily  designed to search for sub-GeV dark-sector particles that interact with the SM ones via light mediators, potentially explaining the origin of dark matter (DM). These theoretically well-motivated and cosmologically viable scenarios are challenging to probe with traditional DM detection methods. The experiment, later called NA64e, effectively combining active target and missing-energy techniques and using 50–100 GeV electron and positron beams, achieved the best sensitivity to dark photon models with  hypercharge kinetic mixing. World-leading constraints on Light DM (LDM) obtained from 2016-2023 runs  have been set  \cite{NA64:2023wbi}, probing a part of the key parameter space that explains the observed dark matter relic abundance.  

\par At the same time, other obtained results demonstrate  that  NA64 exceeded its primary goals and developed  a broader dark sector  program created from the initially proposed NA64e  experiment without compromising the searches for LDM. This research program  provides important inputs for many new physics scenarios, including the $^8Be$ anomaly \cite{NA64:2018lsq}, axion-like particles (ALPs) \cite{NA64:2020qwq,Zhevlakov:2022vio}, electron $g-2$ \cite{NA64:2021xzo}, inelastic dark matter \cite{NA64:2021acr, Mongillo:2023hbs}, $B-L$ \cite{NA64:2022yly}, $L_\mu-L_\tau, Z'$ models \cite{Andreev:2024lps}, and Lepton Flavour Violation (LFV) \cite{Gninenko:2018num,Zhevlakov:2023jzt,Radics2023,Ponten:2024grp}. The experiment also complements underground direct detection, neutrino, beam dump, and high-energy collider searches, providing crucial physics input and motivation for the current and future experimental research program at CERN. 

\par In 2018, NA64 proposed extending its search for higher-mass LDM using the SPS M2 160 GeV muon beam \cite{Gninenko:2640930}. A key intermediate goal was to 
probe the simplest and most predictive $L_\mu-L_\tau, Z'$ model, which explains both the muon $g-2$ anomaly, and thermal DM relic density. Pilot muon runs in 2022 nearly ruled out almost completely the $g-2$ explanation and set first constraints on LDM coupled to the second lepton generation \cite{NA64:2024klw}. In 2022, the first pilot measurement was performed by NA64, using a 100 GeV positron beam to maximally exploit the resonant annihilation channel to search for LDM. The results demonstrate the validity of this approach, identifying the corresponding critical items and determining an appropriate strategy to mitigate them, in view of the planned future full-scale $e^+$ program \cite{Bisio:2887649}. A short 2023 test run with a pion beam \cite{NA64:2024azv}, set stringent limits on invisible $\eta$ and $\eta'$ decays, demonstrating that NA64’s sensitivity to leptophobic dark matter could be significantly enhanced in future runs. This opens up an additional line of research at NA64 for exploring a new class of dark sector models with hadron beams.





\section{NA64-$e^-$: status and prospects}

The experiment, in the original electron mode (NA64e), employs the optimized  100 GeV electron beam from the H4 beamline at the North Area (NA). The beam was designed to transport the electrons with the maximal intensity up to a few $\simeq 10^7$ per SPS spill of 4.8 s in the momentum range between 50 and 150 GeV/c.
The hadron contamination in the electron beam was measured to be $\pi/e^- \lesssim 0.5\%$ \cite{Andreev:2023xmj}. 

NA64 combines the active beam dump technique with the missing energy measurement to search for invisible decays of massive $A'$, produced in the ECAL target (the electromagnetic calorimeter) by the dark Bremsstrahlung reaction $e^-Z \rightarrow e^-ZA'$
(where electrons scatter off a nucleus of charge $Z$) \cite{Gninenko:2013rka, Andreas:2013lya}. After its production, the $A'$ would promptly decay into a pair of LDM candidate particles, $A'\rightarrow \chi\chi$ which would escape the setup undetected leaving missing energy as signature. For this reason, we call these searches {\it invisible}. The parameter space characterized by mixing strengths 10$^{-6} < \epsilon <$ 10$^{-3}$ and masses $m_{A'} $ in the sub-GeV range is the NA64 target: a region where the DM origin can be explained as a thermal freeze-out relic. Missing energy experiments, such as NA64, require a precise knowledge of the incoming beam but also an accurate measurement of the deposited energy from the incoming beam's interaction.

\begin{figure}[t]
\centering
\includegraphics[width=0.8\columnwidth]{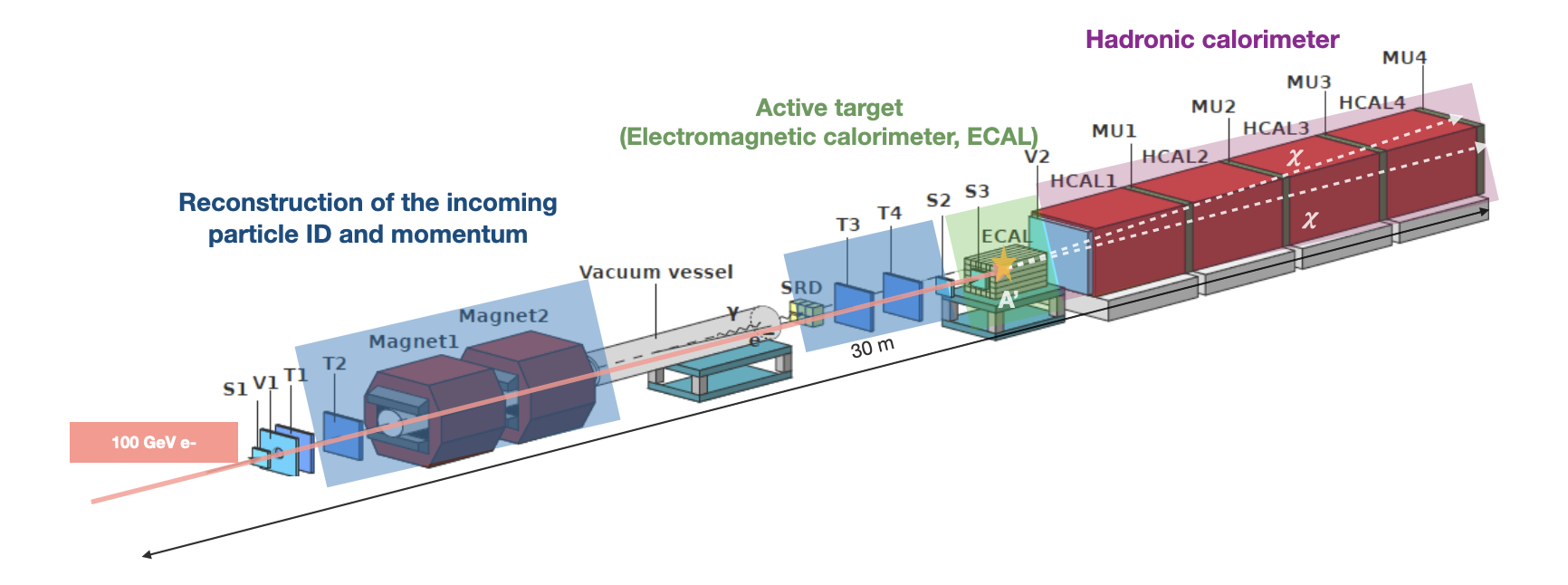}
\caption{\small{NA64 setup and working principle for the search of dark photons through missing energy in the active target (ECAL).}}
\label{fig:NA64_sketch}
\end{figure}

A signal event is identified by a single electromagnetic shower in the ECAL  (with energy $E_{ECAL}$) accompanied by a significant missing energy $E_{miss}=E_{A’}=E_{initial}-E_{ECAL}$.  The occurrence of the $A'$ production is inferred in case these events show an excess above those expected from backgrounds (see Fig. \ref{fig:NA64_sketch} for a sketch of the setup and a summary of the working principle).

\begin{figure}[!b]
\centering
\includegraphics[width=0.45\columnwidth]{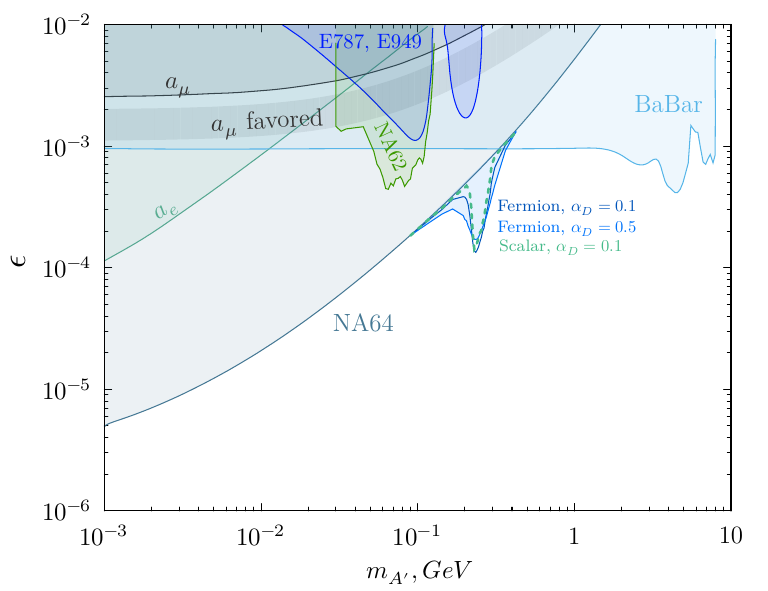}
\includegraphics[width=0.45\columnwidth]{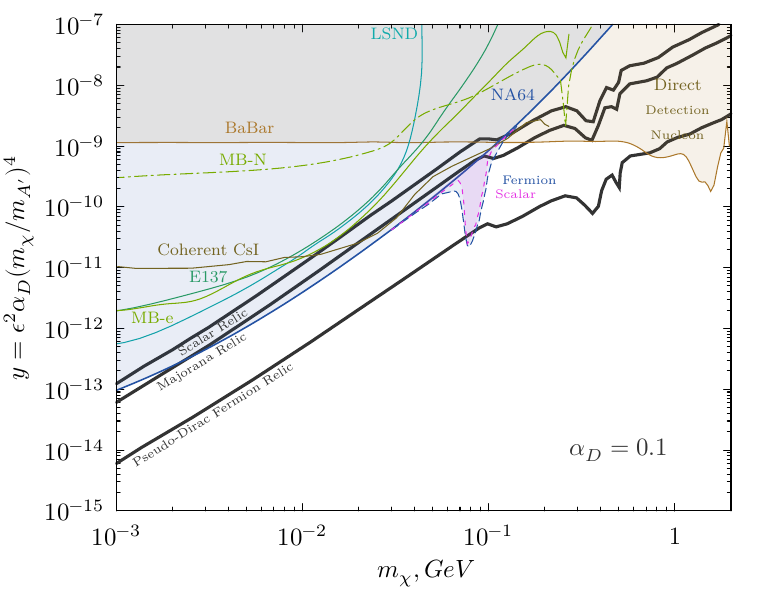}

\caption{\small{Current status of NA64 experiment 90\% C.L. exclusion limits on $A’$ invisible decays, including both the Bremsstrahlung and the resonant $A'$ production channels (Left). LDM searches (Right).  \cite{Andreev:2023xmj}}}.
\label{fig:NA64_invis}
\end{figure}

With the 2016-2022 combined statistics, NA64 sets the most stringent upper limits in the kinetic mixing ($\epsilon$) and dark photon ($A'$) mass plane for masses below 350 MeV. The collected data also allow constraining the values of scalar and Majorana DM with coupling $\alpha_D\leq0.1$ and $m_{A'}>3m_{\chi}$ in the mass range $0.001 \leq m_{\chi} \leq 0.1$ GeV as shown in Fig. \ref{fig:NA64_invis}. The latest results with the 2016-2022 collected statistics were published in Phys. Rev. Lett. \cite{Andreev:2023xmj} and the paper was highlighted as an Editor's suggestion. 

 In addition to the Bremsstrahlung reaction, the resonant $A'$ production channel through the e$^-$ annihilation with the positrons present in the electromagnetic shower has also been considered. The 90$\%$ C.L. exclusion limits from the combined analysis are shown in Fig. \ref{fig:NA64_invis} (the peak present in the limits is the contribution of the resonant annihilation, see more details in Sec. \ref{sec:NA64e+}). The addition of this process to the evaluation of the expected signal yield allows for improvement in the sensitivity in the high mass region, where the yield is suppressed due to the $1/m^2_{A'}$ dependency of the Bremsstrahlung cross-section (see \cite{Marsicano:2018glj} for further details).
 
\begin{figure}[h!]
\centering

\includegraphics[width=\columnwidth]{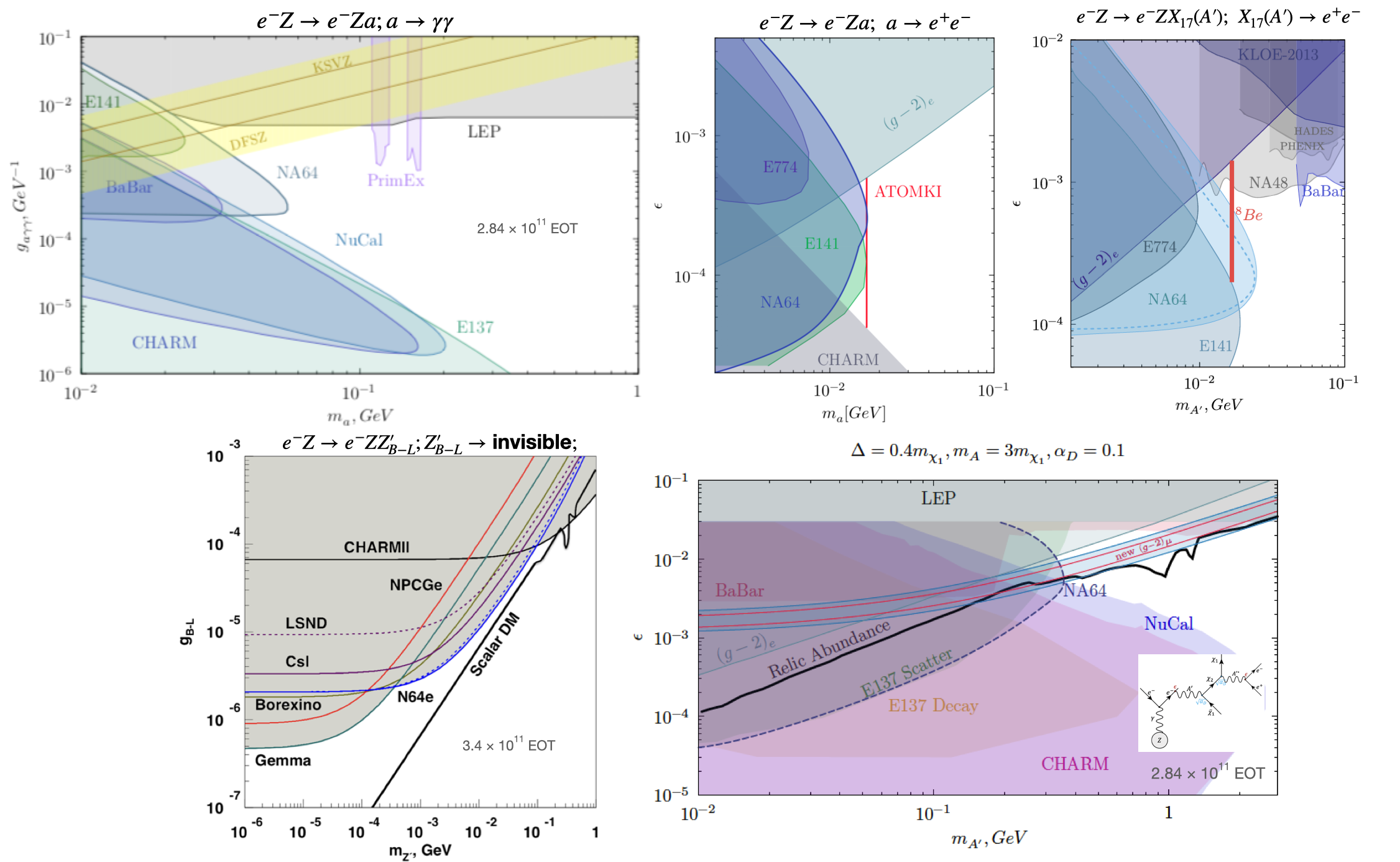}
\caption{\small{Current status of the NA64 experiment 90\% C.L. exclusion limits on ALPs searches \cite{NA64:2020qwq} (top left), pseudoscalar decaying to $e^+e^-$\cite{NA64:2021aiq} (top center) and $A’(X)$ visible decays \cite{Banerjee:2018vgk} (top right), and NA64 coverage for a new B-L Z' gauge boson \cite{NA64:2022yly} (bottom left) and semi-visible $A'$ decays  \cite{NA64:2021acr} (bottom right).}}
\label{fig:NA64_all}
\end{figure}

NA64 has also a great potential to explore scenarios beyond the dark photon hypothesis. NA64 has already demonstrated sensitivity to light-scalar and pseudo-scalar axion-like particles (ALPs) produced via the Primakoff reaction \cite{NA64:2020qwq}, closing part of the gap between beam-dump and LEP bounds (see Fig.~\ref{fig:NA64_all}, top left). The experiment also excluded a significant portion of the parameter space relevant for the hypothesized X17 boson, which could explain the beryllium anomaly \cite{Banerjee:2018vgk,NA64:2021aiq} (see Fig.~\ref{fig:NA64_all}, top center).

A search for a $Z'$ boson associated with (un)broken $B-L$ symmetry in the keV–GeV range using 2021 data \cite{NA64:2022yly} set the most stringent limits for $0.3 \lesssim m_{Z'} \lesssim 100$ MeV, surpassing constraints from neutrino-electron scattering (see Fig.~\ref{fig:NA64_all}, bottom left). NA64 also explored a $Z'$ boson linked to the muon-tau lepton number difference, which could explain the muon g-2 anomaly and DM relic density \cite{Holst:2021lzm}. With 2016-2022 data, the experiment probed the $(g-2)_{\mu}$-favored region for $1~{keV}<m_Z'<2~{MeV}$ \cite{NA64:2022rme,Andreev:2024lps}.

NA64 is also sensitive to inelastic Dark Matter (iDM) scenarios, where two DM species with a small mass splitting lead to unique topologies involving both missing and visible energy \cite{Mongillo:2023hbs,Izaguirre:2017bqb,Mohlabeng:2019}. An initial recast analysis using 2016–2018 data \cite{NA64:2021acr} probed this signature (see Fig.~\ref{fig:NA64_all}, bottom right) and was featured on the cover of European Journal of Physics C 81/10.

Additionally, NA64 enables searches for Lepton Flavour Violation (LFV) in processes like $e^- N \rightarrow \mu^- N Z'$ \cite{Gninenko:2022ttd}. In 2023, a magnet spectrometer was added to identify final-state muons from electron interactions. The first feasibility study based on Monte Carlo simulations optimized the setup for LFV searches and was recently published \cite{Ponten:2024grp}.

 To further improve sensitivity, as part of the upgrades to the setup of NA64 at the CERN SPS H4 beamline, a prototype veto hadron calorimeter (VHCAL) was installed in the downstream region of the experiment during the 2023 run. The VHCAL, made of Cu-Sc layers, was expected to be an efficient veto against upstream electroproduction of large-angle hadrons or photon-nuclear interactions, reducing the background from secondary particles escaping the detector acceptance. With the collected statistics of $4.4\times 10^{11}$ electrons on target (EOT), we demonstrate the effectiveness of this approach by rejecting this background by more than an order of magnitude. This result provides an essential input for designing a full-scale optimized VHCAL to continue running background-free during LHC Run 4, when we expect to collect $10^{13}$ EOT. 

\section{NA64-$e^+$: status and prospects}\label{sec:NA64e+}

The use of a positron in a missing-energy experiment allows to maximally exploit the resonant annihilation channel to search for LDM. In the reaction, a positron from the shower - either the primary or a secondary one - annihilates with an atomic electron, resulting to the production of a LDM pair; in the simpler model, this involves the exchange of an on-shell dark photon ($e^+ e^- \rightarrow \Apr \rightarrow \chi \overline{\chi}$)\cite{Marsicano:2018krp}. Thanks to the resonant nature of the process, the signal yield is strongly enhanced in the allowed kinematic region, roughly corresponding to the interval $\sqrt{2m_eE^{thr}_{miss}} \lesssim m_\Apr \lesssim \sqrt{2m_eE_0}$, where $E_0$ is the beam energy and $E^{thr}_{miss}$ the missing energy threshold. 
The resonant reaction is also characterized by a unique signature of the signal, manifesting itself as a narrow peak in the missing energy distribution, whose position solely depends on the $\Apr$ mass value. 

\begin{figure}[t]
    \centering
    \includegraphics[width=0.48\textwidth]{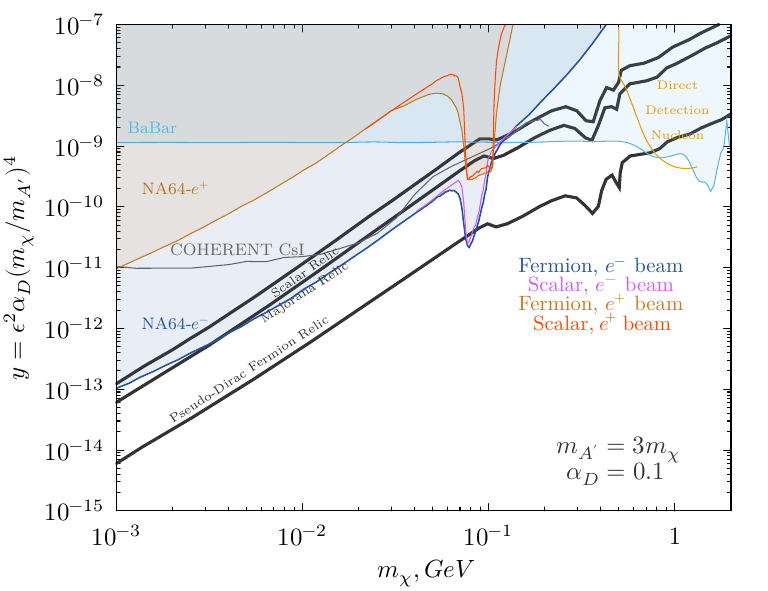}
    \includegraphics[width=0.48\textwidth]{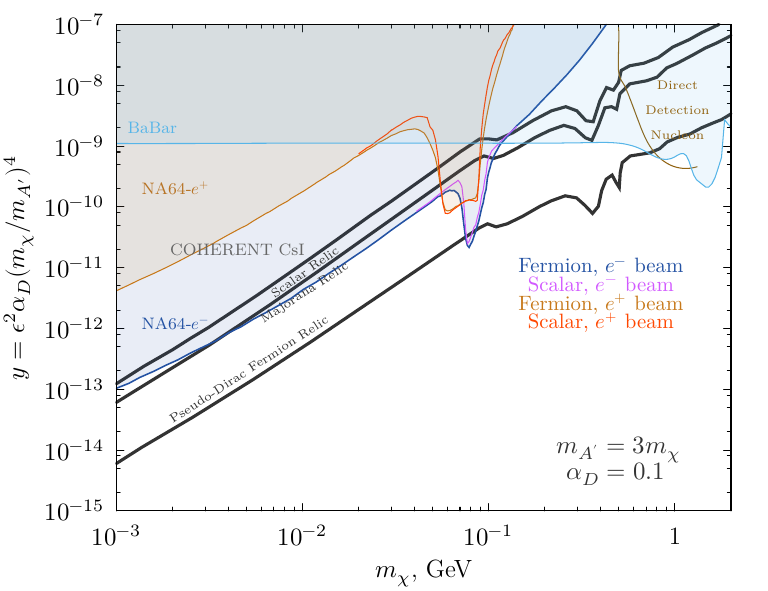}

     \includegraphics[width=0.5\linewidth]{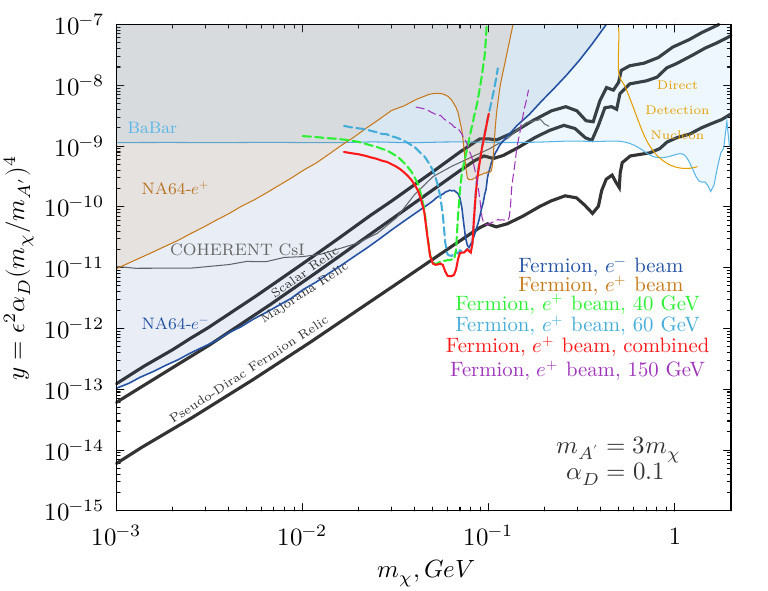}
    \caption{\small{Top-left: the exclusion limits reported by NA64 from the 2022 positron-beam missing energy measurement at 100 GeV, with $10^{10}$ accumulated $e^+$OT. Top-right: same result from the 2023 positron-beam run at 70 GeV, with comparable statistics. Bottom: sensitivity of the post-LS3 NA64 positron-beam program, with two measurements at 40 and 60 GeV beam energy, each with total statistics of about $10^{11}$ $e^+$OT}}
    \label{fig:posi1}
\end{figure}

A first pilot measurement was performed by NA64 in summer 2022, using a 100 GeV positron beam and accumulating a total statistics of $\simeq 10^{10}$~$e^+$OT~\cite{NA64:2023ehh}, with average intensity of about $5\times10^{6}$ particles/spill. The main goal of the measurement was to demonstrate, for the first time, the validity of the positron-beam missing-energy technique to search for LDM, identifying the corresponding critical items and determining an appropriate strategy to mitigate them, in view of a future full-scale $e^+$ program. During this measurement, the H4 beamline optics was reversed to transport positively-charged particles, with a corresponding beam purity of about $96\%$~\cite{Andreev:2023xmj}, slightly larger than the value corresponding to the negative-charge configuration. Due to this, the largest background source expected in the measurement, with an expected yield of about 0.06 events, was from the in-flight decay of misidentified $\pi^+$ and $K^+$ to a $e^+ \nu_{e}$ pair, with the soft electron giving rise to a low-energy EM shower in the ECAL. After scrutinizing the data, no events were observed in the signal region. This allowed NA64 to set new exclusion
limits that, relative to the collected statistics, prove the power of this experimental technique, as depicted in Fig.~\ref{fig:posi1}, left panel.

Motivated by this result, in 2023 NA64 performed a second positron-beam measurement, accumulating a total statistics of $1.6 \times 10^{10}$ $e^+$OT at 70 GeV beam energy. The lower beam energy allowed to probe a different region of the LDM parameter space, by varying the position of the resonant annihilation peak. The main goal of this measurement was to probe the hermeticity of the NA64 detector at lower beam energy, in view of a high-statistics post-LS3 program with multiple beam energies down to 40 GeV.  Thanks to the improved hermeticity of the setup, mostly because of the installation of the VHCAL detector prototype in front of the ECAL, the signal region was extended by setting $E^{thr}_{miss}=28$~GeV. In this configuration, the dominant background source was indeed due to events in which the primary beam particle interacts with upstream detector elements -- scintillator counters and tracking detectors -- resulting to a soft positron hitting the calorimeter and one or more energetic secondary hadrons emitted at large angle, outside the detector acceptance. The corresponding yield, estimated from data by side-band projections, was about 0.09 events. No events were observed in the signal region, and new exclusion limits were set. Thanks to the strong enhancement of the signal yield due to the $e^+e^-$ annihilation mechanism, particularly effective for a primary positron beam, NA64 explored a new region in the LDM parameter space, 
excluding the existence of vector-mediated light dark matter in the mass range $165 < m_\Apr < 220$ MeV, for $\varepsilon$ values down to $2.5\times10^{-4}$ and $\alpha_D=0.1$ (see Fig.~\ref{fig:posi1}, right panel). To further enhance this result and probe other $\alpha_D$ values, in 2024 NA64 performed a third positron run at 70 GeV beam energy, accumulating a comparable statistics. To cope with the reduction of synchrotron-radiation emission at lower beam energy, the setup was improved by replacing the existing Synchrotron Radiation Detector with an optimized LYSO-based compact calorimeter. The data analysis is currently in progress, and results are expected by the end of 2024.

These successful efforts paved the way to the development of the NA64 post-LS3 program with positron beams, presented to the CERN SPSC in 2024~\cite{Bisio:2887649}. This program is supported by a dedicated ERC grant, POKER (``POsitron resonant annihilation into darK mattER''), aiming at a broad LDM search with positron beams, missing energy measurements~\cite{Marini:2024jmj}. Its goal is to perform a comprehensive measurement with positrons, with multiple runs at different energies, to ``scan'' the LDM parameter space for masses larger than $\approx 100$~MeV, by varying the position of the resonant peak. Specifically, in the first phase (2028-2029) we plan to accumulate up to $10^{11}$ $e^+$OT at 60 and 40 GeV, to probe the LDM parameter space $135$~MeV~$\lesssim m_\Apr \lesssim 250$~MeV down to the coupling values predicted by the Pseudo-Dirac Fermion Relic target model, for $\alpha_D=0.1$ (see also Fig.~\ref{fig:posi1}). In the second phase (2030-2032), instead, we aim at enhancing the accumulated statistics to probe larger $\alpha_D$ values, up to $\alpha_D=0.5$, by running the experiment at larger beam intensity, with up to $3\times10^{7}$ positrons per SPS spill. This program presents significant challenges, mostly connected to the hermeticity of the setup at lower beam energy, as well as to the possibility of running the experiment at very high beam intensity. NA64 has already started a dedicated R$\&$D program to tackle them, supported by the results already obtained from the 70 GeV measurement. Specifically, in order to enhance the background rejection capability of the detector at 40 GeV, a large VHCAL detector will be installed in front of the calorimeter. Another item of concern is the fact that, at 40 GeV, penetrating particles produced in the ECAL and interacting in the HCAL with a small energy deposition may be misidentified as zero-HCAL-energy events due to resolution effects of the latter. To mitigate this effect, we plan to improve the detector readout scheme, to increase the light collection efficiency.


\section{NA64-$h$: status and prospects}
The NA64h is a  fixed target experiment utilizing the pion, kaon, and proton beams at the CERN SPS and capable of detecting signal of BSM physics via hadron-nuclear scattering. 
It  employs  the missing energy 
technique  and some of its specific signature  \cite{Gninenko:2014sxa}.  The NA64h  can explore a variety of sub-GeV DM models interacting with the SM via light mediators. Since the dominant production and detection modes for DM occur via hadronic nuclear scattering reactions,  NA64h has a unique sensitivity to leptophobic (or hadrophilic) DM models with a mediator coupled predominantly to light quarks.  These modeles  could yield the correct thermal relic abundance and 
are difficult to test with  NA64e and  NA64$\mu$  experiments. 
\par The NA64h plans to perform a  comprehensive study of  leptophobic models predicting
\begin{enumerate}[(i)]
\item a new $U'(1)$ gauge bosons or scalars coupled to LDM \cite{McElrath:2005bp,Batell:2018fqo}
\item invisible of semi-visible decays of neutral and vector mesons $\eta, \eta', \omega, \rho, ...$ \cite{Gninenko:2014sxa,Zhevlakov:2023wel,Gninenko:2023rbf}, and, in particular,   $K_{S,L} \to invisible$  decays of neutral kaons, which have never been  probed \cite{Gninenko:2014sxa,Gninenko:2015mea, Hostert:2020gou}. The later is highly complementary to searches for $K^{+,0} \to \pi + invisible$ decays. 
\item   $K^0 - K'^0$ oscillations of neutral $K^0$ into its dark partner, e.g.,  in the Mirror Matter model \cite{Gninenko:2024ujp, Gninenko:2025xmb}. This search is complementary to experiments looking for $n - n'$ oscillations. 
\item  a class of dark sector models with a heavy neutral lepton \cite{Bertoni:2014mva} can also be tested.
\end{enumerate}

In 2024, NA64h  published the first results from a proof-of-concept search for Dark Sectors via invisible decays of pseudoscalar $\eta$ and $\eta'$ mesons  at the CERN SPS \cite{NA64:2024azv}. Our novel technique uses the charge-exchange reaction of 50 GeV $\pi^-$ on nuclei of an active target as the source of neutral mesons. The $\eta, \eta' \to invisible$ events would exhibit themselves via a striking signature - the complete disappearance of the incoming beam energy in the detector. No evidence for such events has been found with  $2.9\times10^{9}$ pions on target accumulated during one day of data taking. This allows us to set a stringent limit on the branching ratio ${\rm Br}(\eta' \to invisible) < 2.1 \times 10^{-4}$, improving the current bound by a factor of $\simeq3$. We also set a limit on ${\rm Br}(\eta \to invisible) < 1.1 \times 10^{-4}$ comparable with the existing one. These results demonstrate the great potential of our approach and provide clear guidance on how to enhance and extend the sensitivity for Dark Sector physics for future searches for invisible neutral meson decays \cite{Gninenko:2023rbf}. We are currently considering the possibility of performing this program at the CERN PS T9 beamline. The first 2 weeks of test beam at T9 are foreseen in October 2025. An addendum to the SPSC with our future plans after LS3 is in preparation.

\section{NA64-$\mu$: status and prospects}

At the end of 2021, NA64 began its complementary program for exploring Dark Sectors at the M2 beamline of the CERN SPS. This unique beam can deliver high energy 160 GeV muons with maximal intensities reaching up to $2 \times 10^8$ muons/spill, an order magnitude higher than that available for electrons and positrons at H4. This higher intensity potentially allows for acquiring significantly larger statistics, enhancing the sensitivity of searches for rare processes and feebly interacting particles. New particles ($X$) could be produced in the bremsstrahlung-like reaction of impinging 160 GeV muons on an active target, $\mu N\rightarrow\mu NX$, followed by their decays, $X\rightarrow\text{invisible}$. In this case, the experimental signature would be a scattered single muon from the target, with about less than half of its initial energy and no activity in the sub-detectors located downstream the interaction point \cite{Gninenko:2014pea}. To demonstrate the experiment proof-of-principle we optimised a pilot run in 2022 to carry out such search using a sub-GeV $Z'$ charged under $U(1)_{L_\mu-L_\tau}$ as benchmark process. 

\begin{figure}[h!]
    \centering
    \includegraphics[width=0.47\textwidth]{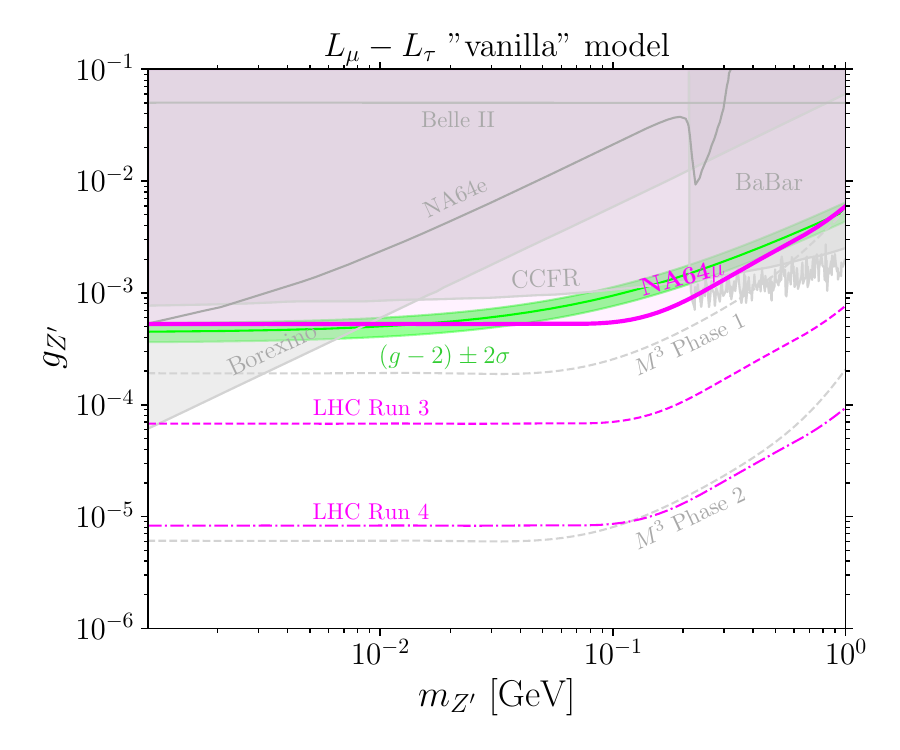}
   \includegraphics[width=0.47\textwidth]{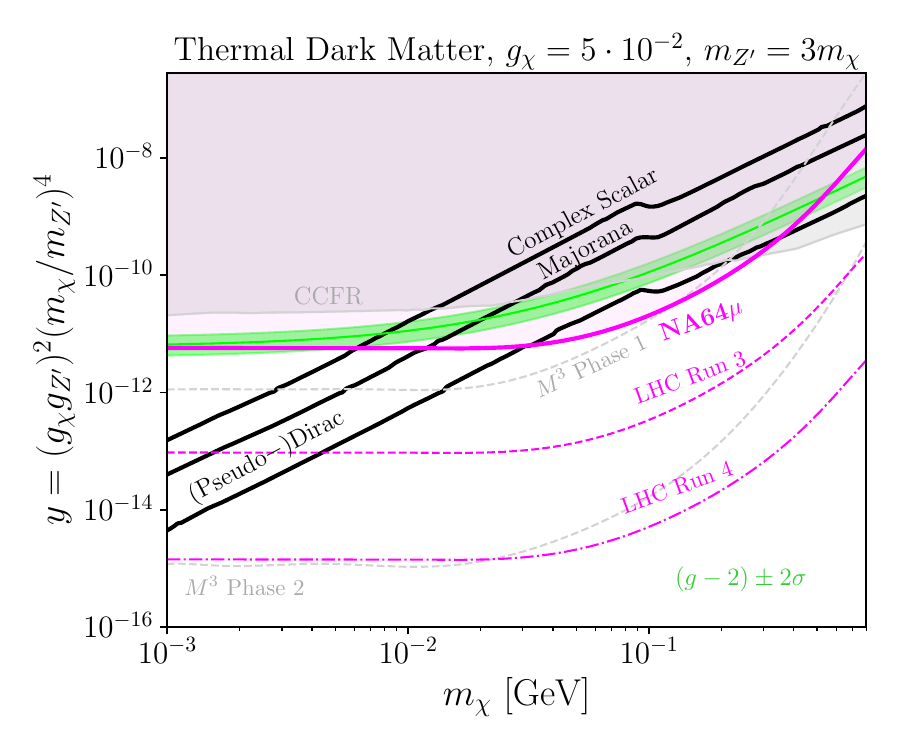}
    \caption{({\it Left}) NA64$\mu$ 90\% CL exclusion limits on the coupling $g_{Z^\prime}$ as a function of the $Z'$ mass, $m_{Z'}$, for the vanilla $L_\mu-L_\tau$ model. 
    ({\it Right}) The 90\% CL exclusion limits obtained by the NA64$\mu$ experiment in the $(m_\chi,\ y)$ parameters space for thermal Dark Matter charged under $U(1)_{L_\mu-L_\tau}$ with $m_{Z'}=3m_\chi$ and the coupling  $g_\chi=5\times10^{-2}$ for $1.98\times10^{10}$ MOT. The branching ratio to invisible final states is assumed to be $\text{Br}(Z'\rightarrow\text{invisible})\simeq1$. 
    Existing constraints from other experiments, and projections before and after LS3 are also shown.}
   \label{fig:2022results}
\end{figure}

For the collected $1.98\times 10^{10}$ MOT, the expected background within the signal box was estimated to be $0.07\pm0.03$. The main background source comes from the momentum mis-reconstruction in the magnet spectrometer downstream of the target. After unblinding, no events compatible with an invisible $Z'$ charged under $U(1)_{L_\mu-L_\tau}$ decay in the signal region were found. These results allowed us to obtain the 90\% confidence level upper bounds on the coupling as a function of the $Z'$ mass. In Fig. \ref{fig:2022results} our constraints for two scenarios are illustrated. The left plot considers the minimal or \emph{vanilla} model where $Z^{'}$ decays only to neutrinos. The green band shows the $\Delta a_\mu$ $\pm2\sigma$ region for the $Z'$ contribution to the $(g-2)_\mu$ anomaly. In this case, NA64$\mu$ limits exclude masses $m_{Z'}\gtrsim40$ MeV and coupling $g_{Z^\prime}\gtrsim6\times10^{-4}$. On the other hand, in extended scenarios, the $Z'$ can decay to DM particles. NA64$\mu$ could probe a portion of the $(m_\chi,\ y)$ parameter space, with $y=(g_\chi g_{Z'})^2(m_\chi/m_{Z'})^4$. For $m_{Z'}=3m_{\chi}$, chosen to be away from the resonant enhancement $m_{Z^\prime}\simeq2m_\chi$, and $g_\chi=5\times10^{-2}$ (to probe the DM freeze-out prediction and the muon g-2 anomaly), our results constrain the dimensionless parameter $y$ to $y\lesssim6\times10^{-12}$ for $m_\chi\lesssim40$ MeV.

\begin{figure}[h!]
    \centering
    \includegraphics[width=0.47\textwidth]{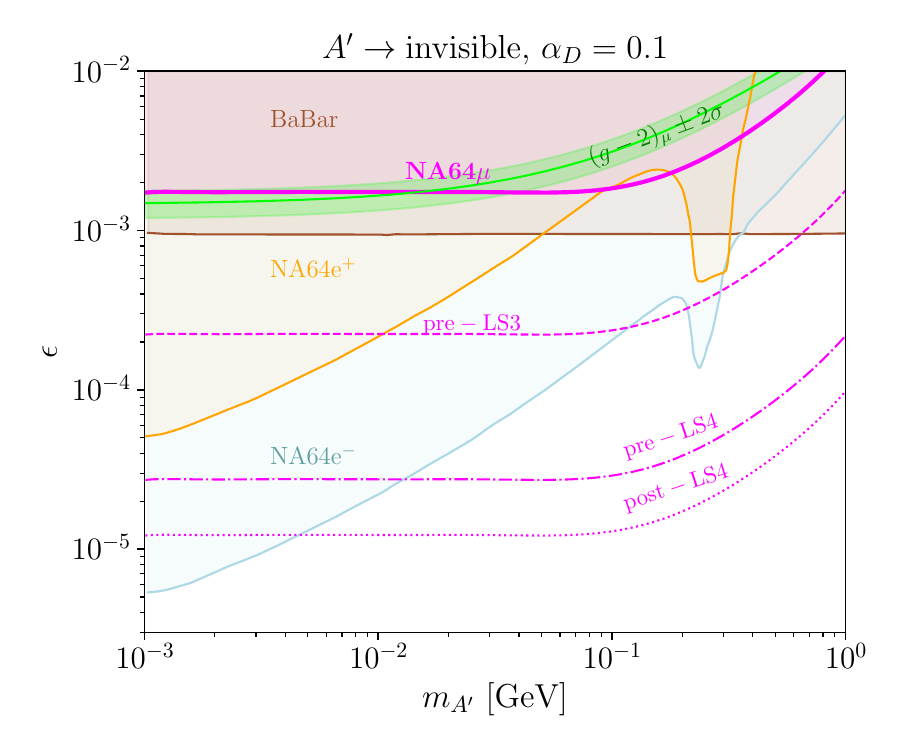}
    \includegraphics[width=0.47\textwidth]{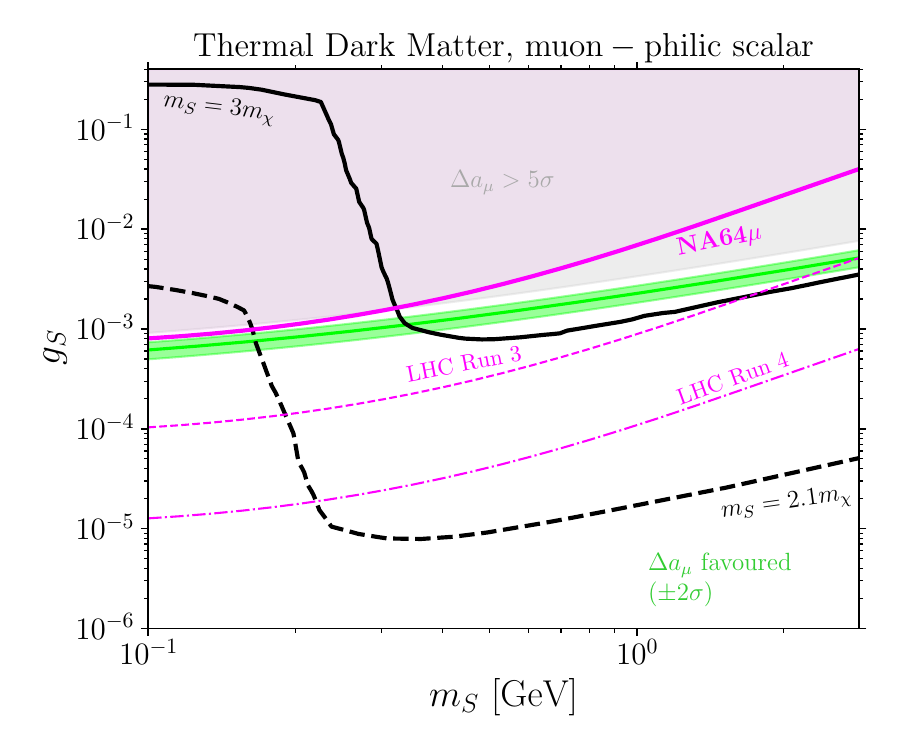}
    \caption{NA64$\mu$  90\% CL excluded limits with $1.98\times10^{10}$ MOT and projected limits for the LHC Runs 3 and 4 for a Dark Photon ({\it left}) and for a muonphilic scalar mediator S ({\it right}) \cite{NA64:2024nwj}. 
    }
    \label{fig:sensitivity-ltdm}
\end{figure}

NA64$\mu$ has also a unique sensitivity to other scenarios complementing the physics case accessible with electron and positron beams. As an example, in Fig. \ref{fig:sensitivity-ltdm}, the 90\% C.L. limits from NA64$\mu$ for scalar and vector mediators are illustrated \cite{NA64:2024nwj}. 
The plot on the left, shows the results for Dark Photon searches using a muon beam. The current limit is not yet competitive compared to the other experiments but demonstrates our capability to probe masses above $100$ MeV with higher statistics as shown with the projections for NA64$\mu$ before and after LS3. As shown in Fig. \ref{fig:sensitivity-ltdm}, combined with our $e^-$ and $e^+$ beam programs NA64 can probe the full parameter space of benchmark LDM models.

In the right plot of  Fig. \ref{fig:sensitivity-ltdm}, the sensitivity in the case of a scalar mediator is depicted for the choice of parameters of coupling $g_\chi=1$. Two different scenarios containing Dirac DM with the benchmark ratio $m_S/m_\chi=3$ and the near-resonant regime $m_S/m_\chi=2.1$ are considered. The corresponding thermal targets are extracted from \cite{Chen:2018vkr}. Our limits cover part of the parameter space compatible with masses $m_S\leq300$ MeV up to a coupling $g_S\sim10^{-3}$ for the mass ratio $m_S/m_\chi=3$. In the case, $m_S/m_\chi=2.1$, the limits are only covering scalar masses up to $m_S\sim m_\mu$. More details on (i) the Monte Carlo (MC) approach used in simulating the signal events, (ii) systematics in the signal yields and (iii) level of background extracted from data can be found in \cite{NA64:2024nwj}. 

 These limits are the first ones obtained using a high-energy muon beam \cite{NA64:2024klw}. These results open a new path to explore DS physics in a complementary way to present and future experiments, highlighting the robustness of our novel missing energy-momentum technique, which is planned to be used by other experiments such as LDMX and M3 \cite{Kahn:2018cqs,Solt:2023igr}. The results have been published in Phys. Rev. Lett. and have been featured in the Physics Magazine of the American Physical Society (APS) \cite{NA64:2024klw}. 
\begin{figure}[h!]
    \centering
    \includegraphics[width=0.47\textwidth]{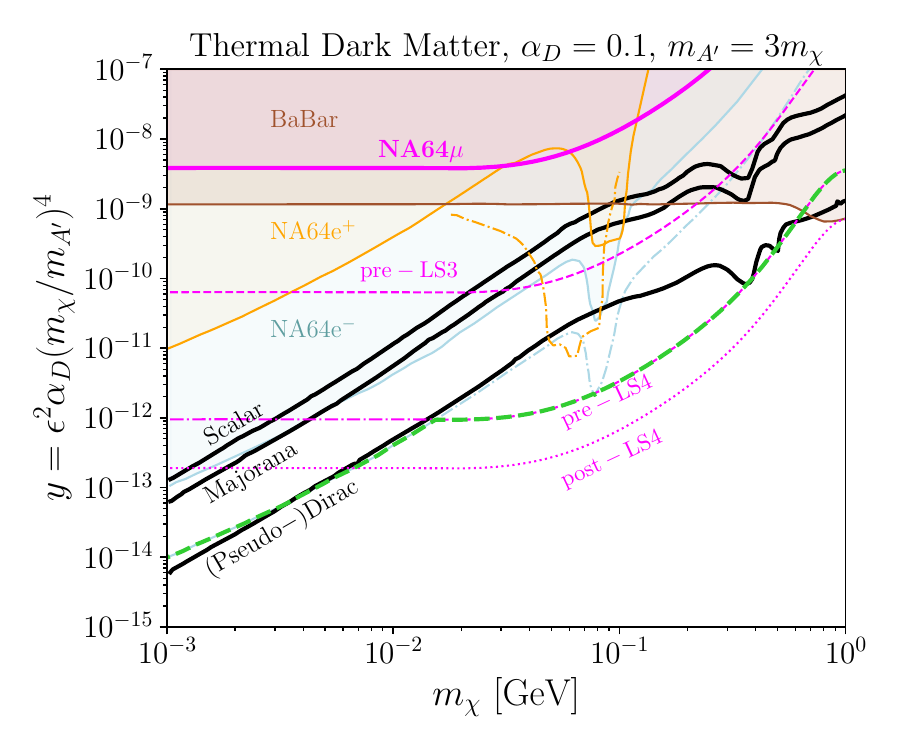}
    \includegraphics[width=0.47\textwidth]{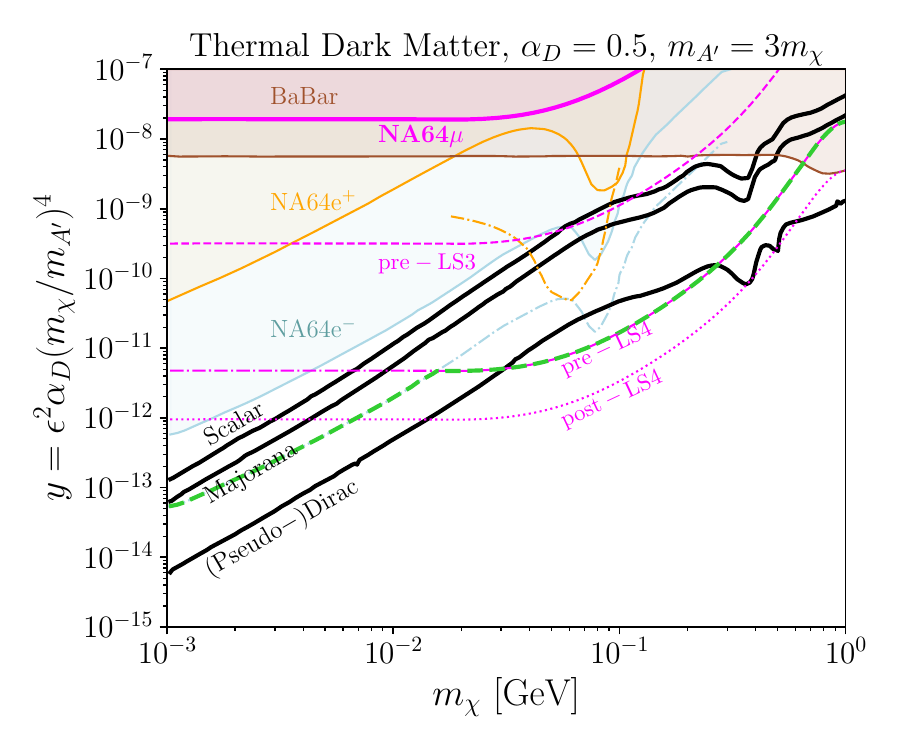}
    \caption{Combined NA64 90\% CL excluded limits with $1.98\times10^{10}$ MOT and projected limits for Light Dark Matter mediated by a Dark Photon. 
    }
    \label{fig:combinedNA64-ltdm}
\end{figure}
\par The projected sensitivities from the muon experimental program before the CERN Long Shutdown (LS3), or \textit{LHC Run 3}, and after LS3, \textit{LHC Run 4} are also shown in Figs \ref{fig:2022results}, \ref{fig:sensitivity-ltdm} and \ref{fig:combinedNA64-ltdm}. For example, as shown in   Fig. \ref{fig:combinedNA64-ltdm}, the full coverage of the parameter space of benchmark LDM models  from combined results obtained with  electron, positron and muon beams is expected.
The computation of those limits is based on the foreseen detectors' upgrade to cope with higher beam intensities and (ii) the background reduction due to improvement of the experimental set-up configuration that was already implemented in 2023-2024 data taking. The total statistic was improved by a factor 20 compared to the 2022 pilot run. The preliminary results are encouraging showing that the main background source arising from momentum miss-reconstruction could be reduced by two orders of magnitude and the tracking efficiency was improved by a factor of four. During LS3 further upgrades will be required to cope with the very high intensity that is potentially available at M2. Those include improving the detector hermeticity, adding new trackers, upgrading the calorimeters and trackers readout and developing a new trigger scheme. Those improvements can also expand our scientific goals and we plan to study our sensitivity to scenarios such as ALPs.


\section{Summary}
The searches for dark sector and dark matter with the fixed-target NA64 experiment at the CERN SPS have almost ten years long  and remarkable history. 
Being approved in 2016 for the
LDM search with the pioneering  active beam dump plus missing energy  techniques at electron beam \cite{NA64:2023wbi}, NA64 since then has developed its application for incisive exploration  of dark sectors  with   positron \cite{NA64:2023ehh}, muon \cite{NA64:2024klw}, and recently, hadron \cite{NA64:2024azv} beams.  The experiment has successfully met its primary objectives, as outlined in the EPPS input (2018), producing results which demonstrate its potential for a high-sensitivity  search for Dark Sectors and LDM and its ability to operate
 in a near-background-free environment. It has also  exceeded its primary goals and, without compromising the searches for LDM,  provides important inputs for other BSM  scenarios.
 The PBC and FIPs communities recognize the high scientific value of NA64’s contributions and its complementarity to other ongoing and planned experiments and support its continued exploration. 

With the improvement of the SPS performance after LS3, we anticipate a great opportunity for the further sensitive  exploration of the dark sector  and other important new physics. 
To fully exploit its physics potential, NA64 will undergo an upgrade during LS3 to continue running in background-free mode at the higher SPS beam rates. Planned upgrades include (a) improved detector hermeticity with a new veto hadron calorimeter, (b) enhanced particle ID with a synchrotron radiation detector, and (c) increased beam rates via upgraded electronics. For NA64h, the best location is under study, including the PS T9 beamline, while true muonium production at H4 is being optimized.
With the recently strengthened NA64 collaboration, stable operations and timely data analysis are planned for LHC Run 4. 
\par The  sensitivity of $\lesssim 10^{-13}$ for 
the dark sector searches, expected to be achieved  with $\sim 10^{13}$ electrons, $\sim 10^{11}$ positrons (40 and 60 GeV), and $\sim 2\times10^{13}$ muons on target, will allow NA64 to explore new parameter space of LDM and many other well-motivated models,  with the potential for its discovery or conclusive exclusion. 

\bibliographystyle{unsrt}
\bibliography{bibl}

\end{document}